\documentclass[aps,prd,preprintnumbers,showpacs]{revtex4}
\usepackage[dvips]{graphicx}
\begin{document}

%
%

\eprint{Nisho-1-2021}
\title{Spectral-temporal features of repeating ( one-off ) FRBs and Axion Star }
\author{Aiichi Iwazaki}
\affiliation{Nishogakusha University,\\ 
6-16 Sanbantyo Chiyoda Tokyo 102-8336, Japan.}   
\date{Apr. 25, 2021}
\begin{abstract}
The fast radio bursts ( FRBs ) are energetic radio bursts with millisecond duration only observed at radio frequencies.
The generation mechanism is still mysterious. 
We have proposed a generation mechanism of both repeating and one-off FRBs.
They arise from axion star collision with neutron star or magnetized accretion disk of galactic black hole.  
Once we accept the existence of the axions, we find that
the mechanism well explain previously observed spectral-temporal features.  In this paper
we show that it also explains recently observed 
phenomena such as downward drifting in the repeating FRBs, etc..
Analysis of the downward drifting based on Doppler effects has been presented
in recent papers, in which a superradiance system of molecular or atom has been proposed as a source of FRBs.
We apply the analysis to our mechanism and find that 
it well explains the relation between the downward drifting rate and the duration  
of the repeating FRBs. The Doppler effects lead to the fact
that the duration of radio burst with higher center frequency is shorter than that of radio burst 
with lower center frequency in the repeating FRBs.
Our generation mechanism naturally explain polarization angle swing
observed in the repeating FRB180301 and one-off FRBs. 
We also discuss the association between the FRB200428 and magnetar SGR J1935+2154.
The X ray burst observed just after the observation of the FRB 
could be triggered by the axion star collision with the magnetar. 
We also explain the consistency of our generation mechanism with observed 
spectral-temporal differences in the repeating and one-off FRBs, e.g. 
longer duration ( smaller luminosity ) of repeating FRBs than duration ( luminosity ) of one-off FRBs.

\end{abstract}
\hspace*{0.3cm}
\pacs{ Axion, Fast Radio Burst, Accretion disk}

\hspace*{1cm}

\maketitle

\section{introduction}
Fast radio bursts\cite{review} are energetic radio emissions with millisecond duration and are considered to be 
extragalactic events. 
Hosts of some FRBs have been identified and shown to be extragalactic. They show a variety of morphology. 
Since the first discovery\cite{frb} of fast radio burst ( FRB ), more than a hundred FRB sources have been detected\cite{rep,rep1,rev}.
We assume that they are categorized into two types; repeating FRB and one-off FRB.
Their sources are essentially different.
The repeating FRBs\cite{repeat,repeat0,repeat1} emit radio bursts repeatedly so that their spectral-temporal features have been obtained in detail.
On the other hand, such features in detail of the one-off FRBs have not been observed,
although their observed source number is much more than that of the repeating FRBs. 
There are some
apparently one-off FRBs showing typical features e.g. downdrifting feature of repeating FRBs.
Those could be the candidates for repeating FRBs.
Although we assume two types of FRBs, 
there is a possibility that one-off FRBs 
emit bursts again with long silent period. But, there is an evidence\cite{evolution,different} which shows that their origins are different.  

A number of models\cite{model} for their sources have been proposed, but, there is not yet definite observational evidence
which makes us choose a valid model among them. FRB200428 has recently been observed to be associated with magnetor SGR 1935+2154. 
Radio and X ray bursts have been detected\cite{coincidence,coincidence2,diffecoincidence} simultaneously. Although it strongly suggests that 
magnetor is a source of the FRBs,
it is possible that a generation mechnism of the FRB such as axion star collision with the magnetor, becomes the ignition of the X ray burst.
In these circumstances,
the spectral-temporal features observed in the repeating FRBs are fairly effective to distinguish the models. 
Similarly, the difference in the features of repeating and one-off FRBs is also important factor to identify real generation mechanism.
The real generation mechanism of the FRBs must be consistent with any observational results.

\vspace{0.1cm}
The spectral-temporal features of the repeating FRBs, which we consider significant in this paper, are in the following.
First, it appears that the repeating FRBs are narrow band\cite{frequency,bandwidth,bandwidth1,bandwidth2,narrowband} 
and their bandwidths $\delta\nu $ are proportional to 
their center frequencies $\nu_c$\cite{narrowband,narrowband1}; roughly, $\delta\nu \sim 0.16\nu_c$.
The repeating FRBs have been detected in the frequency range such as $120$MHz $\sim 8$GHz\cite{frequency,frequencyL,frequency2L}. 
The center frequencies $\nu_c$ are in the band range.
But, they have not yet been detected in much lower or higher frequency than those in the range.
The duration $t_W$ of radio burst with higher center frequency is shorter than that of radio burst 
with lower center frequency in the repeating FRBs\cite{frequency,duration}.  
 
Secondly, some of them are composed of several sub bursts and the sub bursts show
so called downward drifting\cite{drifting}. That is, the radio wave with higher frequency arrives earlier than 
the radio wave with lower frequency,
even when we correct dispersion measure effect. 
In particular, it is remarkable that a relation\cite{drift-duration} between downward drift rate and duration of the bursts
excellently holds even for different repeating FRBs. The relation has been pointed out in references\cite{drift-duration}, in which
authors present a model of the downward drifting. 
As we explain later,
the feature arises due to Doppler effects 
by the relativistic motions of the sources for the repeating FRBs.

Thirdly, although
it appears that the emissions of the repeating FRBs randomly take place, two cases of periodicity have been observed;
$\sim16$ days for FRB180916.J0158+65\cite{periodicity18} and $\sim 157$ days for FRB121102\cite{periodicity12}. In particular,
the periodicity of the FRB180916.J0158+65 is characterized such that there are active period of about
$4$ days within which the emissions of the radio bursts actively occur. Such active periods 
are periodically repeated with the period $\sim16$ days. It is considered that the periodic actvities arise due to 
periodic motions\cite{periodic} of astrophysical objects emitting the repeating FRBs, e.g. axion star rotating galactic black hole. 

\vspace{0.1cm}
Furthermore, it is important to notice that there is difference or similarity of spectral temporal features between the repeating 
and one-off FRBs. The difference or similarity is also useful to identify the real generation mechanism of the FRBs.
In general, the fluences of the repeating FRBs observed 
with WRST/Apertif and CHIME/FRB\cite{resmallflux} are smaller than those of one-off FRBs
observed with CRAFT\cite{nonsmallflux}. Furthermore,
the duration of the repeating FRBs is larger on average than that of the one-off FRBs, when we compare them
at the identical frequency $\sim 1.4$GHz; see also the reference\cite{effelsberg121102}.
 In addition, there are a number of data growing that
the repeating FRBs have narrow bandwidths, while 
such narrow bandwidths have not been observed in one-off FRBs. 
Probably, the bandwidths of one-off FRBs are much larger than the observational band range of current radio telescopes.
According to our model, these differences arise from the different strengths of magnetic fields possessed by the sources emitting
the repeating and one-off FRBs.
  
Some of the repeating FRBs have been observed to show a significant fraction of linear polarization\cite{polarization} 
and a variety of polarization angle swing\cite{reswingpalarization} .
The swing of the polarization angle has also been observed in one-off FRBs 
such as FRB181112\cite{swingpolarization} and FRB110523\cite{nonswingpolarization}.
The polarization swing is often observed in pulsars. It 
suggests that the radio waves of the FRBs pass the plasma involving strong magnetic field changing rapidly its direction
along the line of sight. 
These different or similar features of FRBs are useful to distinguish generation models of the FRBs.
The real generation mechanism must explain all of these features mentioned above. 

\vspace{0.1cm}
We have previously proposed a generation mechanism\cite{axionmodel,axionmodel1} of FRBs using axion star.
( Different generation mechanisms of FRBs using axion star and axion minicluster have been proposed\cite{others}. )
The QCD axion\cite{axion} is one of the most promising candidate of dark matter.
Simultaneously, it solves strong CP problem.
The axion star\cite{axionstar} is gravitationally bounded state of such dark matter axions.
The axions are formed in early universe and may occupy a large fraction of the dark matter. 
In particular, some of them may condense\cite{minicluster,axionstar} and form axion stars.
According to our generation mechanism\cite{axionmodel,narrowband},
the one-off FRBs arise from the collision between the axion star and neutron star, while the repeating FRBs do from the collision
between the axion star and magnetized accretion disk around galactic black hole. 
We have explained most of the spectral-temporal features mentioned above with the mechanism.
For instance, those are the presence of a variety of center frequencies $120$MHz $\sim 8$GHz, finite bandwidth in the repeating FRBs and the differences
of fluence and duration between repeating and one-off FRBs. The features has been further confirmed by recent observation since
we presented our previous papers\cite{axionmodel,narrowband}.
But, recent observations such as the downward drifting, the periodicity of the repeating FRBs, 
polarization angle swing e.t.c. have not yet been addressed in our previous papers. 
In this paper we would like to explain such newly observed phenomena as well as the features noticed previously.

\vspace{0.1cm}
For the purpose, we should notice
recent papers\cite{drift-duration} in which a spectral-temporal structure of the repeating FRBs is
well explained by a simple model based on Doppler effect.
In particular, a relation between the downward drifting rate and duration of the bursts
has been shown to be remarkably well satisfied observationally. 
Their papers have analyzed the FRBs using superradiance\cite{superradiance} from molecules or atoms as
a generation mechanism of the bursts. The point is that the emitter of the FRBs moves
with relativistic velocity relative to the observer.

\vspace{0.1cm}
In this paper,
we briefly explain the essence of the references in section \ref{doppler} and phenomena of downward drifting. In order to apply it to our generation mechanism
of repeating FRBs and one-off FRBs in the section \ref{feature}, we
briefly sketch our mechanism in section {\ref{axionmodel}} for clarity to understand the essence of our model.	 Then, we
explain QCD axion and axion star in section \ref{axion}. 
We proceed to discuss in detail our generation mechanism of FRBs in section \ref{axionfrb}.
Our generation mechanism of the repeating FRBs is axion star collision with magnetized accretion disk of galactic black hole.
The velocity of the disk from which the FRBs are emitted, can be relativistic. 
In this section we present the update version of previous estimation\cite{narrowband,axionmodel}
of the luminosity in the FRBs. We have included the effects of tidal forces when the axion star collides accretion disk of black hole or neutron star itself.
In section \ref{feature}, we apply the essence in the references
to our axion star model for the repeating FRBs. We show that Doppler effects due to the accretion disk cause
the downward drifting mentioned above. As a consequence of the analysis, 
we obtain the typical strength ( $\sim 10^{10}$G ) of the magnetic field in the disk. 
The strength is a hundred times weaker than that of neutron star. 
 In section \ref{difference}, we show that the difference in strength of magnetic fields
of accretion disk ( $\sim 10^{10}$G ) and neutron star ( $\sim 10^{12}$ ) G causes 
different spectral features of the repeating and one-off FRBs such as duration, luminosity, etc.. 
In section \ref{6}, we show the consistency of our model with recent observations 
such as FRB's association with magnetor, polarization angle swing, etc..
In section \ref{7}, we discuss that we can determine the axion mass with the observation of the bandwidth ( and localization of host galaxy ) of the one-off FRBs.
In the final section \ref{8},  after summarizing our results, we discuss\cite{newmethod} a new method for 
detecting dark matter axion using superconductor and existing radio telescope.

\vspace{0.1cm}
In our discussions below there are several review sections in which we explain our previous results and idea presented in the reference \cite{drift-duration}.
The main sections for presenting new results are sections \ref{axionfrb}, \ref{feature}, \ref{difference} and \ref{6}.  


\section{a model of downward drifting in spectrum }
\label{doppler}
We will explain the model\cite{drift-duration} for the downward drift observed in repeating FRBs.
It is supposed in the model that the source of the FRBs moves with velocity $\vec{V}$ respect to the observer ( the earth ). 
The observed frequency $\nu_{obs}$ of the radio wave emitted by the source 
receives Doppler effect such that $\nu_{obs}=\nu_0\Pi(V,\theta)$ with $\Pi(V,\theta)=\sqrt{1-V^2}/(1-V\cos\theta)$ where $\nu_0$ denotes the 
frequency of the wave measured at the rest frame of the source, and $\theta$ does the angle 
between the velocity $\vec{V}$ of the source 
and the line of sight. Similarly, the time scale $\tau_0$ measured at the rest frame is shifted such that 
$\tau_{obs}=\tau_0\Pi(V,\theta)^{-1}$ at the frame of the observer.

Burst is emitted by the source when the source is stimulated. But there is a delay time between the emission and the stimulation.
For example, dipole radiation is emitted by an electron when harmonically oscillating electric field is imposed on the electron.
As long as the strength of the electric field is so weak as for the harmonic motion of the electron to be thermally disturbed
( or for the electron not to emit sufficiently strong radiation to be observable ),
the electron cannot emit the dipole radiation.  But as the electric field gradually increases, 
the electron begins to emit observable dipole radiation.
In this way, the delay time between the emission and the stimulation e.g onset of the electric field is present. 

\vspace{0.1cm}
The model in the reference describes a time sequence of the burst emission. First a trigger happens to stimulate the source at $t'=0$.
Then, the emission happens at $t'=\tau'_D$ after the stimulation. The emission lasts for $\tau'_W$.
The time scales are measured at the rest frame of the source.
On the other hand, when the time scales are measured at the observer, the delay $t_D$ and the duration $t_W$ 
are given such that $t_D=\tau'_D\Pi(V,\theta)^{-1}$
and $t_W=\tau'_W\Pi(V,\theta)^{-1}$.

Therefore we obtain important relations in the model,

\begin{equation}
\label{D}
t_D=\tau'_D\frac{\nu_0}{\nu_{obs}} \quad \mbox{and} \quad t_W=\tau'_W\frac{\nu_0}{\nu_{obs}}.
\end{equation}  

Obviously, we cannot observe the delay time $t_D$. 
But we can observe the difference $\delta t_D$ of the delay times between two radio waves.
This is the difference of the arrival times between two radio waves. 

Using the relation in eq(\ref{D}), the authors of the references derive a relation between downward drifting rate $\delta \nu_{obs}/\delta t_D$
and the duration $t_W$ of the bursts,

\begin{equation}
\label{formula}
\frac{\delta \nu_{obs}}{\delta t_D}=-\frac{\nu_{obs} A}{t_W} 
\end{equation}
with  $A=\tau'_W/\tau'_D$, where we note the relation $\delta t_D=-\tau'_D\delta \nu_{obs} \nu_0/\nu_{obs}^2$ derived from eq(\ref{D}).

The formula in eq(\ref{formula}) has been shown to be remarkably well fitted to the observations\cite{drift-duration}.
Here we explain the drifting rate $\delta \nu_{obs}/\delta t_D$ which characterizes the temporal structures of repeating FRBs.
Some of the bursts in the repeating FRBs
are composed of a few or several sub-bursts. Each of the sub-bursts shows downward drifting with increasing arrival time.
That is, within a sub-burst, the radio wave with higher frequency arrives faster
than the radio wave with lower frequency arrives, 
even if we correct the effect of dispersion measure. 
The above formula describes the fact that the radio wave with the lower frequency $\nu_{obs}+\delta \nu_{obs}$ ( $\delta\nu_{obs} <0$ ) 
arrives at the time $t=\delta t_D$ later after the arrival of the radio wave with the higher frequency $\nu_{obs}$. The delay time $t=\delta t_D$
is proportional to the duration $t_W$ of the sub-burst; $\delta t_D=-t_W\delta\nu_{obs}/(\nu_{obs}A) $.
The coefficient $A$ is observationally determined.
Obviously the formula predicts the presence of bursts with upward drifting with increasing arrival time, i.e. $\delta \nu_{obs} >0$.

\vspace{0.1cm}
It should be stressed that the formula has been shown to hold for sub-bursts over wide range of frequencies $\nu_{obs}$.
That is, it holds for the bursts with center frequency $\nu_c \sim 300$MHz to bursts with $\nu_c \sim 8$GHz.
( Multiwavelength radio observations\cite{bandwidth1} suggest 
that each of sub-burst has a finite bandwidth $\Delta\nu$, which is proportional to the center frequency $\nu_c$ of the burst. 
Roughly, $\Delta\nu\simeq 0.16 \nu_c$\cite{narrowband, narrowband1}. So, the bursts are specified by their center frequencies $\nu_c$. )
Remarkably, the coefficient $A\sim 0.08$ is almost identical to all of the bursts, although their origins of the sources may be different. 
They are
FRB121102, FRB180916.J0158+65 and FRB180814.J0422+73, respectively.
In any ways, the essence of the model is in the formula in eq(\ref{D}). It is derived by taking account of Doppler effect
which is caused by the velocity of the source to the observer.
As we show below, 
the sources of the repeating FRBs move with relativistic velocity respect to the observer.
The idea explained here is very general and can be applicable to any generation mechanism of the repeating FRBs.
We apply it to our axion star model.

\section{axion star model of FRBs}
\label{axionmodel}
In order to roughly catch a picture of our generation mechanism for FRBs, 
here we briefly explain the essence of our axion model for the repeating and one-off FRBs, respectively. 
Our generation mechanism\cite{bandwidth, narrowband}
for the repeating FRBs is in the following. They arise from the collision between strongly magnetized accretion disk
and axion star. The disk is assumed to be geometrically thin and would be 
present around galactic black hole with mass of the order of $10^3M_{\odot}\sim 10^5M_{\odot}$.
Electrons in the disk are forced to oscillate with oscillating electric field, which is 
induced by the axion star under the strong magnetic field of the disk.
That is, such an oscillating electric field is generated when the axion star is under the magnetic field.  
As is well known, 
an axion with small momentum is converted into a photon with the momentum under homogeneous magnetic field. 
Furthermore, the axion star is a gravitationally loosely bounded state of axions. It is
composed of coherent axions with small momenta of the order of the inverse of axion star radius $R_a\sim 100$km. 
Thus, the energies of the coherent axions forming the axion star are equal to the axion mass. They
are converted into
coherent photons with the small momentum, which represent the oscillating electric field 
with the frequency given by the axion mass $m_a$; $m_a =10^{-6}\rm eV\sim 10^{-3}$eV window allowded for QCD axion. 
The oscillation is harmonic with the frequency $m_a/2\pi$
because the momentum of the axion is extremely smaller than the mass $m_a$.

The electrons forced by the electric field emit dipole radiations with the frequency $m_a/2\pi$.
Because the oscillation of the electrons is coherent over the 
spatial range $R_a^2\delta$ with skin depth $\delta$ up to which emission arises without absorption, a large amount of
coherent dipole radiations are emitted. Because we assume sharp boundary between vacuum and surface of
accretion disk, the skin depth can be defined similar to that in metal.
( In the actual collision, the axion star is distorted\cite{tidalforce} by 
tidal force and its form becomes long stick with width $r_a$ of the order of  $0.1$km. Thus, the spatial range of coherence is
of the order of $r_a^2\delta$. )

Such axion stars with large number would be 
present as a dark matter around galactic black hole, and
frequently fall into the black hole. When they collide with the magnetized geometrically thin accretion disk
around the black hole, the FRBs repeatedly arise.
This is our generation mechanism of the repeating FRBs. 
We should stress that the velocity of the accretion disk can be relativistic near the black hole.
The frequencies of the repeating FRBs are Doppler shifted. The observed spectral-temporal features 
can be well understood with
the model, as we explain below in detail. 

On the other hand, one-off FRBs arise from the axion star collision with neutron star in our model\cite{axionmodel}.
The radiation mechanism is the same as that in the magnetized accretion disk.
The main difference between the accretion disk and neutron star is the strength of magnetic field. 
The magnetic field of the neutron star is much stronger than that of the accretion
disk of galactic black hole. Furthermore, electron gas in the disk can rotate with relativistic velocity, 
while the one in the neutron star rotates with non relativistic velocity.
These differences lead to the difference in spectral-temporal structures of the one-off FRBs
and the repeating FRBs, as discussed in section (\ref{difference})

\section{axion and axion star}
\label{axion}
We explain dark matter axion and axion star.
The axion is a promising candidate of dark matter in the universe. 
The axion is the Nambu-Goldstone boson\cite{axion} associated with U(1) Pecci-Quinn symmetry. 
The symmetry is chiral and naturally solves the strong CP problem in QCD. The axion is called as QCD axion.
Although the axion is a real massless Nambu-Goldstone boson, it acquires 
its mass $m_a$ through chiral anomaly because the Pecci-Quinn symmetry is chiral.
The anomaly gives rise to axion potential, i.e. axion self-interactions.
The mass $m_a$ of the axion is approximately restricted in the range of $10^{-6}$eV $\sim 10^{-4}$eV by cosmological
models\cite{cosmology} and lattice gauge theories\cite{lattice}.
In this paper we consider the QCD axion and assume that 
the dark matter is composed of the axions and that some of them may form the axion stars. 
( There are models of axion-like particles which interact with electromagnetic fields in the similar way to the QCD axion.
Their masses are not restricted in such a small range and may take values in extremely broad range. 
Here, we only consider the QCD axion, which is supposed to be more realistic than the axion-like particles. )

\vspace{0.1cm}
The axion star\cite{axionstar} which is mainly considered in this paper is gravitationally bound state of axions.
The axion stars are characterized by two parameters, the mass $M_a$ of the axion star
and the axion mass $m_a$. 
When the mass $M_a$ of the axion star is small enough
for the binding energies of the axions to much less than the axion mass $m_a$ ( so the energy of the axion is $\omega_a\simeq m_a$ ),  
spherical form of the axion star ( the derivation of the solution is presented in our paper\cite{axionstar} ) is approximately given by

\begin{equation}
\label{a}
a(t,r)=a_0f_a\exp(-\frac{r}{R_a})\cos(m_a t) \quad \mbox{with} 
\quad a_0=3.2\times 10^{-7}\Bigl(\frac{M_a}{4\times 10^{-12}M_{\odot}}\Bigr)^2\Big(\frac{m_a}{0.6\times10^{-5}\mbox{eV}}\Big)^3
\end{equation}
with the decay constant $f_a$ of the axions and radial coordinate $r=|\vec{x}|$,   
where the radius $R_a$ of the axion stars is approximately given by
\begin{equation}
R_a=\frac{1}{GM_a m_a^2}\simeq 
180\mbox{km}\Bigr(\frac{0.6\times10^{-5}\mbox{eV}}{m_a}\Bigl)^2\frac{4\times10^{-12}M_{\odot}}{M_a}
\end{equation}
where $G$ denotes the gravitational constant.
The decay constant $f_a$ is related with the mass $m_a$; 
$m_a\simeq 6\times 10^{-6}\mbox{eV}\times (10^{12}\mbox{GeV}/f_a)$.  
The order of the mass $M_a\sim 10^{-12}M_{\odot}$ ( or radius $R_a\sim 100$km ) of the axion star 
was obtained\cite{axionmodel,rate} by comparing the collision rate between neutron stars
and axion stars in a galaxy with the event rate of one-off FRBs $\sim 10^{-3}$ per year in a galaxy.
If the actual event rate is more large than $\sim 10^{-3}$ per year in a galaxy,
$M_a$ is smaller than $10^{-12}M_{\odot}$.
In the estimation, the axion star was assumed to occupy a large fraction of the dark matter.
( In our model, one-off FRB arises from the collision between the axion star and neutron star. )
It was found that the mass $M_a$ takes a value roughly given by $M_a=(10^{-11}\sim 10^{-12})M_{\odot}$.
Obviously the axions composing the axion star are coherent because the number of axions 
in the volume $m_a^{-3}$ is huge $(M_a/m_a)/(R_a m_a)^3\sim 10^{41}$. 
( Although the collision rate of the axion star and the sun is extremely small,   
if it collides the sun, it simply passes through the sun without any observable radio emission
because magnetic field $\sim 1$G in photosphere of the sun is
too weak for radiation to be observed. )

We should mention that the important feature of the axion star is the Harmonic oscillation of the
axion star $a(t)\propto \cos(m_at)$. The oscillation leads to the Harmonic oscillating electric field
when the axion star is under external magnetic field $B$. The strength of the electric field
is proportional to $B$ so that its influence on strongly magnetized electron gas such as neutron star
is very important. Namely, it generates FRBs.

\vspace{0.1cm}
In the actual collision, the axion star is distorted\cite{tidalforce} by 
tidal force of black hole or neutron star. Its form becomes long stick 
with width $r_a$ of the order of  $0.1$km; $r_a\propto R_a$.
Thus, we should note that the spatial range of coherence when it collides magnetized electron gas in accretion disk of black hole
or neutron star, is
of the order of $r_a^2\delta$.

\vspace{0.1cm}
The axion star is called as dilute axion star, which is gravitationally loosely bounded state. 
It is characterized with the small amplitude of the axion field i.e. $a/f_a \sim a_0 \ll 1$.
It is a solution of axion's free field equation coupled with gravity. The potential of the axion field $a(\vec{x},t)$ plays no important role
except for the mass term.
On the other hand, dense axion star with $a_0 \sim 1$ is present, which is a state bounded by the axion potential.  
Gravitational force plays no important role. The radius of the dense axion star is six or more order of magnitude less than that of the 
dilute axion star. For instance $R_a\sim 1$m for $M_a\sim 10^{-12}M_{\odot}$ and $m_a\simeq 6\times 10^{-5}$eV.
The dense axion star is more hardly affected by tidal force around neutron star or black hole than the 
dilute axion star. Although they are physically interesting, they are not stable\cite{axionstar} against the emission of axions.
Thus, even if they are formed in the early universe, they would decay within cosmological time scales and are not observable.

\section{generation mechanism of repeating fast radio burst}
\label{axionfrb}
When the axion star collides the magnetized accretion disk, the electric field is induced by the magnetic field $\vec{B}$ in the disk, 

\begin{eqnarray}
\label{elec}
\vec{E}_a(r,t)&=&-\alpha \frac{a(\vec{x},t)\vec{B}(\vec{x})}{f_a\pi}
=-\alpha \frac{a_0\exp(-r/R_a)\cos(m_at)\vec{B}(\vec{x})}{\pi} \nonumber \\
&\simeq& 5.2\times 10^{-1}\,\mbox{eV}^2( \,\,\simeq0.7\times 10^4\mbox{eV}/\mbox{cm} \,\,)\cos(m_at)
\Big(\frac{M_a}{4\times 10^{-12}M_{\odot}}\Big)^2\Big(\frac{m_a}{0.6\times10^{-5}\mbox{eV}}\Big)^3\frac{B}{10^{10}\mbox{G}}\frac{\vec{B}}{B}.
\end{eqnarray} 
at $r < R_a$ and with the fine structure constant $\alpha\simeq 1/137$,
where we have used the solution $a(\vec{x},t)$ in eq(\ref{a}) representing the axion star. 
We have also taken into account that the momenta $\sim 1/R_a$ of the axions are vanishingly small; $\vec{\partial}a(\vec{x},t)\simeq 0$.
We suppose that the strength of the magnetic field $B$ in the accretion disk\cite{mag} is of the order of $10^{10}$G.
We note that the electric field oscillates with frequency $m_a/2\pi$ because it is generated by the axion star $a(t)\propto \cos(m_at)$.

The electric field is induced by the magnetic field because the axion $a(\vec{x},t)$ couples with electromagnetic fields in the following,

\begin{equation}
\label{L}
L_{aEB}=k_a\alpha \frac{a(\vec{x},t)\vec{E}\cdot\vec{B}}{f_a\pi}
\end{equation}   
where the numerical constant $k_a$ depends on axion model.
The standard notation $g_{a\gamma\gamma}$ is such that $g_{a\gamma\gamma}=k_a\alpha/f_a\pi\simeq 0.14(m_a/\rm GeV^2)$
for DFSZ model\cite{dfsz} and $g_{a\gamma\gamma}\simeq -0.39(m_a/\rm GeV^2)$ for KSVZ model\cite{ksvz}.
In other words, $k_a\simeq 0.37$ for DFSZ and $k_a\simeq -0.96$ for KSVZ.
In eq(\ref{elec}) we have set $k_a=1$. The coupling shows that electric field $\vec{E}$ is generated by the source term $a(\vec{x},t)\vec{B}$.
The coherent axions $a(\vec{x},t)$ are converted into the coherent photons $E(\vec{x},t)$ under the magnetic field $B$.
( We can easily derive\cite{newmethod} the electric field in eq(\ref{elec}) 
by solving a modified Maxwell's equations involving the effect of the axion photon interaction in eq(\ref{L}). ) 

\vspace{0.1cm}
The electric field $\vec{E}=\vec{E}_0(r)\cos(m_at)$
make electrons in the disk oscillate with the frequency $m_a/2\pi$.
Actually, Electron's equation of motion is effectively given such that $\vec{\dot{p}}=e\vec{E}-\tau^{-1}\vec{p}$
with relaxation time $\tau$.
Then, the solution is $p\sim eE_0\tau\cos(m_at)$ because $\tau \ll 1/m_a$. 
Then, the electric current is given by $en_eE_0\cos(m_at)/m_e$ with electron mass $m_e$ and number density $n_e$.
Thus, when the edge ( $r>R_a$ ) of the action star collides the edge of the accretion disk, the electric field $E(r>R_a)$ is 
too small for radiation emitted by the oscillating electrons to be observable.
But the amount of the radiation rapidly increases with the increase of the electric field $E(r<R_a)$. 
In other words, at a time $t=0$ ( trigger ) just when the edge of the axion star collides the edge of the disk, 
the electric field $E_0(r\gg R_a)$ at the edge of the disk is very weak. No observable radio emission arises.
But, the electric field $E_0(r)$ rapidly increases as $r\to 0$. ( The center of the axion star approaches the edge of the disk. )
Then, the observable radio emission begins at $t=\tau'_D$. 

The above equation of motion is the effective one taken on average over the
individual collisions among electrons. The momentum of each electron may be composed of two terms, 
the oscillation term $p_{os}=p_0\cos(m_at)$ and the thermal motion $p_{th}$; $\langle p_{th}\rangle =0$.
Intuitively, when the thermal fluctuation $p_{th}$ is larger than the oscillation,
the dipole radiation is disturbed. On the other hand, when the oscillation $p_{os}$ is larger than $p_{th}$,
the radiation arises. The time when the oscillation overcomes the thermal fluctuation is just $\tau_D'$.

\vspace{0.1cm}
The Harmonic oscillation of the electrons is coherent over the spatial range beyond the Compton wave length $(m_a/2\pi)^{-1}$ 
because the electric field and the magnetic field $B$ of the disk
are uniform over the length. The coherent length ( inverse of its momentum ) 
of the axion star when it collides with accretion disk or neutron star is
of the order of $r_a\sim 0.1$km. ( The axion star is distorted by tidal force to the form of long stick with width $r_a\propto R_a$. )
Thus, the dipole radiation\cite{narrowband} from the electrons in the accretion disk is coherent and sufficiently strong to explain
observed luminosity of repeating FRBs,

\begin{eqnarray}
\label{W}
\dot{W}&=&\dot{w}(n_er_a^2\delta)^2 \sim 10^{46}\,\mbox{erg/s}\,\Big(\frac{n_e}{10^{20}\mbox{cm}^{-3}}\Big)\Big(\frac{eB}{10^{10}\mbox{G}}\Big)^2\Big(\frac{m_a}{0.6\times10^{-5}\mbox{eV}}\Big)^{-3} \\
\dot{w}&=&\frac{e^4E_0^2}{3m_e^2}\sim 10^{-11}\mbox{erg/s}\Big(\frac{M_a}{4\times 10^{-12}M_{\odot}}\Big)^4
\Big(\frac{m_a}{0.6\times 10^{-5}\mbox{eV}}\Big)^6\Big(\frac{B}{10^{10}\rm G}\Big)^2
\end{eqnarray}
with the radiation power of each electron $\dot{w}$,
where the number density $n_e$ of electrons ( magnetic field $eB$ ) is assumed to be $10^{20}\rm cm^{-3}$ ( $10^{10}$G ).
We also assume that the temperature $T$ of electron gas is $10^6$K; see the comment just below. 
The dipole radiations are linearly polarized as observed\cite{polarization}.
The present estimation of the luminosity is the elaborated update version of the previous estimations\cite{narrowband,axionmodel}. 

\vspace{0.1cm}
Here we should make a comment on the magnitude of the luminosity.
The emission arises in the coherent region with the volume $r_a^2\delta$ in the edge of the accretion disk
where $\delta$
is the skin depth ( $\delta=\sqrt{2/\omega\sigma}$ with $\omega=m_a$ and electrical conductivity $\sigma$ ), 
for instance, $\delta\sim 10^{-4}$cm for copper with $n_e\sim 10^{22}$cm$^{-3}$
in room temperature. The electrical conductivity is proportional to 
the electron density $n_e$ and relaxation time $\tau$; $\sigma=e^2n_e\tau/m_e$. Additionally,
the relaxation time due to the electron interactions roughly behaves as
$\tau \propto 1/T^2$ with the temperature $T$. ( The dependence of $\tau$ on $n_e$ is weak; $\tau\propto n_e^{-/3}$. ) 
Thus, $\delta\sim 10$cm for $T=10^6$K and $n_e=10^{20}\rm cm^{-3}$.

\vspace{0.1cm}
In the estimation, we simplify the accretion disk such that it has sharp boundary ( edge ) like metal between vacuum and 
the surface of the disk. Radiations emitted from the disk can reach the earth without absorption. 
The point is that because the radiations from electrons are coherent,
the luminosity is proportional to the large number $(n_er_a^2\delta)^2$.
For instance, $(n_er_a^2\delta)^2\sim 10^{58}$ with $r_s=10^4$cm, $n_e=10^{20}\rm cm^{-3}$ and $\delta=10$cm.
We roughly obtain the above luminosity in eq(\ref{W}), in order to see
whether or not we can have sufficiently large luminosity to meet the observation.

 
Because the precise physical parameters ( number density of electrons, temperature, etc. ) of the accretion disk are unknown, 
the above rough estimation is satisfactory to 
give reasonable values of the real FRB fluxes.
We refer the reference\cite{narrowband} for the previous derivation of the formula
where we have not taken into account the effect of the tidal force. So the previous formula
does not involve the term of $r_a^2\delta$ in eq(\ref{W}).

\vspace{0.1cm}
The oscillation energy is rapidly thermalized in the relaxation time $\tau$ because the electrons collide with each other. 
Then, the temperature $T$ of the electron gas rapidly increases 
so that the conductivity also rapidly decreases; $\sigma=e^2n_e\tau/m_e$ because $ \tau \propto 1/T^2$. 
It implies that the radiation rapidly decreases because of rapid decrease of the oscillating electric current $J=\sigma E_0\propto 1/T^2$. 
Because the temperature increases by a few order of magnitude,
the radiation decreases by the several order of the magnitude.
( The thermalization of the energies of the oscillating electrons is a cause for the termination of radio emission. 
The luminosity rapidly decreases with the thermaliztion. So,
it determines the duration $\tau'_W$. The cause for the termination of radio emission is the one operated for one-off FRBs.
There is another cause for the termination in repeating FRBs as we explain in the next section. )

Thus, we can state that the radio emission stops ( rapid decreases ) in practice 
when the temperature arrives at the oscillation energy $(eE_0(r=0)/m_a)^2/2m_e$.
Intuitively, the thermal fluctuations disturb the coherent oscillations.
Such a critical temperature $T_c$ is given such that
$T_c=(eE_0(r=0)/m_a)^2/2m_e$. Numerically it reads to

\begin{equation}
\label{tem}
T_c=\Big(\frac{eE_0}{m_a}\Big)^2\frac{1}{2m_e}\simeq 0.8\times 10^4\mbox{eV}\Big(\frac{eB}{10^{10}\rm G}\Big)^2
\Big(\frac{M_a}{4\times 10^{-12}M_{\odot}}\Big)^4\Big(\frac{m_a}{0.6\times 10^{-5}\rm eV}\Big)^4 \sim 10^8K.
\end{equation} 

Roughly speaking, the radio emission lasts until the temperature of the electron gas reaches $T_c$.
Therefore,  the radiation from the electrons approximately shows the spectrum such as

\begin{equation}
\label{flux}
S(\nu)\propto \exp\Big(-\frac{(\nu-\frac{m_a}{2\pi})^2}{2(\Delta\nu)^2}\Big),
\end{equation}
with the bandwidths $\Delta\nu=(m_a/2\pi)\sqrt{T_c/m_e}$,
where $\nu$ denotes frequency of radiations. Namely, the radiations are emitted from electron gas with the temperature $T_c$.
The line spectrum of the dipole radiation is 
thermally broadened.  Electron moving toward to the observer emits radiation with higher frequency $\nu>m_a/2\pi$, while
electron moving backward emits radiation with lower frequency $\nu<m_a/2\pi$ at the frame of the observer.
Further, electron gas in the accretion disk moves with the rotation of the disk. Then, the rotation causes Doppler shift of
the frequency $m_a/2\pi$; $\nu_c=\Pi(V,\theta)m_a/2\pi$.
In this way, we find that the line spectrum of the repeating FRBs is broadened by the thermal effect 
and that there is the various center frequency $\nu_c\Pi(V,\theta)m_a/2\pi$, depending on the velocity $V$ of the accretion disk.

\vspace{0.1cm}
Until now, we only consider that the critical temperature $T_c$ is much lower than the electron mass $m_e$. Approximately,
it holds for magnetic field $eB$ 
less than $10^{11}$G. As we show below in next section \ref{feature}, the magnetic fields relevant to the repeating FRBs 
satisfy the condition. That is, examining the observational relation $\Delta\nu\simeq 0.16\nu_c$ 
with the use of $\Delta\nu=m_a/2\pi\sqrt{T_c/m_e}$ and $T_c$ in eq(\ref{tem}), 
we find that the magnetic fields in the accretion disk of galactic black hole is of the order of $10^{10}$G.
When $T_c$ is much less than $m_e$,
the motions of electrons are non relativistic. On the other hand, neutron stars possess strong magnetic field $eB\sim 10^{12}$G.
Then, because $T_c$ is larger than $m_e$, the motions are relativistic.
The critical temperature is determined such as $T_c=\sqrt{p_c^2+m_e^2}\simeq p_c$ with $p_c=eE_0/\sqrt{2}m_a$. ( Here the time average 
$\langle \sin^2(m_a t)\rangle =1/2$ is taken. ) Then, the form of the spectrum from electrons with high temperature $T_c>m_e$
is more complex\cite{dopplerbroaden}
than that of the spectrum eq(\ref{flux}) from non relativistic electrons.
But we can see\cite{dopplerbroaden} that the spectrum of the radiations has peak frequency $\nu_c$ identical to the one of non relativistic electrons,
i.e. $\nu_c=m_a/2\pi$. Similarly, the bandwidth of the spectrum is proportional to the peak frequency; $\Delta\nu \propto \nu_c\sqrt{T_c/m_e}$.  
Such a spectrum arises in the axion star collision with neutron star. One-off FRBs are those emitted
by electrons with high temperature $T_c>m_a$.

\section{spectral-temporal structure of repeating FRB}
\label{feature}
Now, we apply the idea \cite{drift-duration} in section (\ref{doppler}) to our generation mechanism of the repeating FRBs.

The trigger ( stimulation ) happens at $t=0$ when the edge of the axion star collides the surface of the accretion disk.
At the time the radiations are not yet emitted because the oscillating electric field produced in the collision is weak at the edge. 
Then, as the center of the axion star approaches at the surface, the electric field increases and the emission of
the radiation with frequency $\nu$ begins at $t=\tau'_D$ at the rest frame of
the source, i.e. electron gas. 
( The time $\tau'_D$ is determined by the velocity of axion star respect to the disk. The velocity is relativistic 
when it collides because the axion star falls into the location near black hole. That is, roughly $\tau'_D\sim R_a/c$ with light velocity $c$. ) 
The electron gas in the accretion disk moves toward or backward to the observer. Thus,
the delay $\tau'_D$ is shifted such that the delay time at the frame of the observer is given 
such as $t_D=\Pi^{-1}(V, \theta) \tau'_D=\tau'_D\nu/\nu_{obs}$
with Doppler factor $\Pi(V,\theta)=\nu_{obs}/\nu$. $\nu_{obs}$ denotes the frequency at the rest frame of the observer. 
The velocity $V$ relative to the observer 
is that of the electron gas 
in the accretion disk, i.e. the velocity of the accretion disk. It depends on the location in the disk.
The delay has been observed
as a downward drifting in a sub-burst as we have mentioned above.   That is, 
the delay time $t=\delta t_D$ of the radio burst with the frequency $\nu_{obs}+\delta \nu_{obs}$ ( $\delta\nu_{obs} <0$ ) 
after the arrival time $t=0$ 
of the radio burst with the frequency $\nu_{obs}$ is proportional to the duration $t_W$ of the sub-burst; $\delta t_D=-t_W\delta\nu_{obs}/(\nu_{obs}A) $
or $\frac{\delta \nu_{obs}}{\delta t_D}=-\frac{\nu_{obs} A}{t_W}$ with $A=\tau'_W/\tau'_D$. Here we have 
noticed that the duration $\tau'_W$ of the burst is Lorentz contracted such as $t_W=\Pi^{-1}\tau'_W$ at the frame of the observer. 
( The duration $\tau'_W$ is given by the time for which the location the axion star firstly collide moves away from the axion star due to the
rotation of the disk. The direction of the rotation is almost perpendicular to the direction of the velocity of the axion star.
We note that the axion star is deformed by tidal force to a long stick with width $r_s$ whose head moves to the direction
almost perpendicular to the direction of the rotating disk.   
Although the axion star collision with the disk still continues after the time $\tau'_W$, the radiation emitted from the location of the 
continuing collision
would be absorbed by the dense gas ejected by the first collision. )
The axion star moves with relativistic velocity respect to the electron gas. So it is roughly equal to $r_s/c$.  Hence we find that 
$A=\tau'_W/\tau'_D\sim r_s/R_s$ is roughly independent of each FRB events. 
Furthermore, we can explain why the observed value $A=\tau'_W/\tau'_D\sim 0.08$
is smaller than $1$. 
The formula relating the downward drifting rate $\frac{\delta \nu_{obs}}{\delta t_D}$ with the duration $t_W$
has been shown\cite{drift-duration} to be remarkably well fitted to the observations.
Therefore, we find that the downward drifting phenomena within a sub-burst can be well explained in our generation mechanism of the repeating FRBs.
Obviously the formula predicts the presence of bursts with upward drifting with increasing arrival time, i.e. $\delta \nu_{obs} >0$. 

We see that the Doppler effect plays important roles in the downward drifting.
It is caused by the motion of the accretion disk around galactic black hole. 
( We refer papers\cite{magnetormodel} in which the phenomena of the downward drifting is explained using neutron star model 
for repeating FRBs. The model is one of generation mechanisms of FRBs
in standard astrophysics without assuming hypothetical particles such as axion. ) 
A possibility is pointed out that Doppler effect leads to slow radio bursts\cite{slow} emitted from magnetor associated with FRB200428.

\vspace{0.1cm}
We would like to make a comment that there are two causes for the termination of observed radio emission in repeating FRBs. One is the cause
explained above. That is, the termination occurs because the source of the radiation goes away the axion star.
Another cause is the termination due to the thermalization of the oscillating electron energy discussed in the section (\ref{axionfrb}).
The observation indicates that the relaxation time 
of the energies is longer than the duration $\tau'_W$ discussed above. Thus, the duration of repeating FRBs is in practice 
determined by the time $\tau'_W$.

\vspace{0.1cm}
The duration $\tau'_W\sim r_a/c$ of the repeating FRBs is almost independent of the events as we discussed above.
Then we can predict that the observed duration $t_W(\nu)=\Pi^{-1}\tau'_W$ of radiation with frequency $\nu$ 
is shorter than duration $t_W(\nu_0)$ of radiation with lower frequency $\nu_0$,

\begin{equation}
t_W(\nu)=\frac{\nu_0}{\nu} t_W(\nu_0)
\end{equation}
The tendency on average can be seen \cite{frequency}. 

\vspace{0.1cm}
The electron gas in the accretion disk rotates around black hole.  
Thus, the velocity of the disk makes the frequency of the radiation Doppler shifted.  
Then, the frequency $\nu$ of the radiation at the rest frame of the disk is Doppler shifted such that
$\nu_{obs}=\Pi(V,\theta)\nu$ at the frame of the observer, where $V$ is the velocity of the accretion disk at the frame of the observer
and $\theta$ denotes the angle between the direction of $\vec{V}$ and the line of sight. 
The spectrum of the radiations is thermal broadening and is affected by Doppler shifted.
Therefore, the observed spectrum of the radiation is given by

\begin{equation}
\label{obsS}
S_{obs}(\nu)\propto \exp\Big(-\frac{(\nu-\nu_c)^2}{2(\Delta\nu_{obs})^2}\Big),
\end{equation}
with the center frequency $\nu_c=\Pi(V,\theta)m_a/2\pi$ and the bandwidth $\Delta\nu_{obs}=\nu_c\sqrt{T_c/m_e}$ is given by

\begin{equation}
\label{repwidth} 
\Delta\nu_{obs}(\mbox{repeat})=\nu_c\sqrt{\frac{T_c}{m_e}}\simeq 0.13\nu_c\Big(\frac{eB}{10^{10}\rm G}\Big)
\Big(\frac{M_a}{4\times 10^{-12}M_{\odot}}\Big)^2\Big(\frac{m_a}{0.6\times 10^{-5}\rm eV}\Big)^2,
\end{equation}
where $\Delta\nu_{obs}(\mbox{repeat})$ denotes the bandwidth of the repeating FRBs.

When the axion star collide the accretion disk, the spherical form of the axion star is distorted by the tidal force of the black hole
so that it becomes a long stick ( or slender stream ) \cite{axionmodel1, tidalforce} 
with width $r_a$ less than $1$km. The different point in the disk which the axion star collides has 
different velocity $V$, because the accretion disk differentially rotates around the black hole.
 Thus, there are various center frequencies $\nu_c=\Pi(V,\theta)m_a/2\pi=200$MHz $\sim 8$GHz in the repeating FRBs, 
as the previous observations have shown.

\vspace{0.1cm}
Furthermore, the bandwidth $\Delta\nu_{obs}$ is proportional to the center frequency $\nu_c$.
The observations show that roughly $\Delta\nu_{obs}=\nu_c\sqrt{T_c/m_e}\sim 0.16\nu_c$. Using the observational
relation,
we can derive the strength of magnetic field $B$ in the accretion disk relevant to the repeating FRBs.
That is, the relation implies that
the velocity $v_e$ of electrons emitting the bursts
is $\sim 0.2c$ with the light velocity $c$, because $\sqrt{T_c/m_e}=v_e/\sqrt{2}\sim 0.16$.
Then, we can estimate the order of magnitude of magnetic fields in the disk 
using the formula in eq(\ref{repwidth}). 
We find that the magnetic field in the accretion disk relevant to repeating FRBs is given

\begin{equation}
eB \sim 10^{10} \mbox{G}\Big(\frac{M_a}{4\times 10^{-12}M_{\odot}}\Big)^{-2}\Big(\frac{m_a}{0.6\times 10^{-5}\rm eV}\Big)^{-2},
\end{equation}
The order of the magnitude is 
coincident with the strength necessary to
explain total flux $\sim 10^{44}$erg/s of the bursts in eq(\ref{W}).
In this way we can well understand the phenomena observed in the repeating FRBs, the downward drifting and
the narrowband emissions with various center frequencies, $200$MHz $\sim 8$GHz.

\vspace{0.1cm}
The Doppler effect leads to the decrease or increase of the observed flux density,

\begin{equation}
F_{obs}\propto \Pi(V,\theta)^3F_0=\Big(\frac{\nu_{obs}}{\nu_0}\Big)^3F_0
\end{equation}
where $F_0$ denotes the flux density at the rest flame of the source.
It apparently seems that the flux density of radiation with higher frequency is larger than the one of radiation with lower frequency.
But, we note that the radiation with higher center frequency $\nu_{obs}$ is emitted from the location
closer to the black hole than the location from which radiation with lower frequency is emitted.
The velocity of the accretion disk is larger, as its location is closer to the black hole. 
As the axion star is closer to the black hole, the deformation of the axion star by tidal force
is larger. That is, the width $r_a$ of the long stick of the axion star is smaller, as the axion star is closer to the 
black hole. Thus, the coherent region of the emission $r_a^2\delta$ becomes smaller.
Then, the amount of the radiation becomes smaller; it is proportional to $(r_a^2\delta)^2$. Therefore, it is not clear that 
the flux density of the radiation with higher frequency $\nu_c$ is larger than the one of
radiation with the lower frequency.

\vspace{0.1cm}
A burst with very narrow bandwidth $\simeq 65$MHz has recently been observed\cite{bandwidth2} in the repeating FRB190711.
The luminosity of the burst is more than $10$ times smaller compared with other fluence observed in the FRB190711.
The features are related with each other. That is,
according to our generation mechanism,
both features arise from the identical origin, namely, 
the weaker magnetic field in accretion disk than the value quoted above. 
The weak magnetic field $B$ leads to the small luminosity and the narrow bandwidth
of the burst.

\section{difference between one-off FRB and repeating FRB}
\label{difference}

Until now, we have discussed the properties of the repeating FRBs using the axion star model for the generation of the bursts.
To examine the model furthermore, 
we compare spectral-temporal features between the one-off FRBs and the repeating FRBs. 
The differences in the features are caused by the different physical conditions in each sources of the FRBs.
According to our model, the one-off FRBs arise from the axion star collision with neutron star.
The electron gas in the atmosphere of neutron star emits the burst in the collision. 
It is easy to see that the whole of the axion star evaporates with one times collision because it can not pass the neutron star.
Thus, the axion star collision with neutron star generates one-off FRB. 
( The strong magnetic field $B>10^{12}$G induces 
strong electric field $E_a\sim \alpha a(r,t) B/f_a>10^6$eV/cm which generates large amount of electric currents
inside the neutron star with high number density $n_e\gg 10^{23}$ of electrons. The energy of the current is rapidly dissipated.
Namely, the axion star with mass $M_a\sim 10^{-11}M_{\odot}$ ( $\sim 10^{43}$erg ) evaporates inside the neutron star. )

\vspace{0.1cm}
First we note that the repeating FRBs show finite bandwidths, while the finite bandwidths of the one-off FRBs
have not been observed. 
Possibly, the one-off FRBs could not be broadband, but 
their bandwidths could be so large to be over the extent of the frequency ranges of current radio telescopes.
We interpret that it is caused by the difference of the magnetic field in the accretion disk and neutron star. 
Because the magnetic field $eB(\mbox{neutron})\sim 10^{12}$G of the neutron star is much stronger 
than the magnetic field $eB(\mbox{disk})\sim 10^{10}$G in the accretion disk estimated above,
the bandwidth $\Delta\nu_{obs}(\mbox{oneoff}) \propto eB$ of the one-off FRBs is roughly 
a hundred times larger than the bandwidth $\Delta\nu_{obs}(\rm repeat)$ of the repeating FRBs,

\begin{equation} 
\mbox{Ratio of bandwidth}; \quad \quad  \frac{\Delta\nu_{obs}(\mbox{oneoff})}{\Delta\nu_{obs}(\mbox{repeat})}
=\sqrt{\frac{T_c(\mbox{oneoff})}{T_c(\mbox{repeat})}}=
100\times \Big(\frac{eB(\mbox{neutron})}{10^{12}\rm G}/\frac{eB(\mbox{disk})}{10^{10}\rm G}\Big)
\end{equation}
with the bandwidth $\Delta\nu_{obs}(\rm repeat)$ observed at the center frequency $\nu_c=m_a/2\pi$,
where $\nu_c=m_a/2\pi$ is the intrinsic frequency emitted by the electron gas in neutron star.
( We have estimated $m_a/2\pi\sim 1.7$GHz in the previous paper\cite{narrowband} using the data in the observation\cite{axionmass}. )
The bandwidth of the one-off FRBs has not yet been observed.
The bandwidths of the one-off FRBs are over the extent of the frequency ranges of current radio telescopes.
We predict that the bursts showing upper or lower limit of their frequencies are repeating FRBs when they are
observed in the range of the frequency $400$MHz to $2$GHz.   

It is important to notice that the bandwidth $\Delta \nu_{obs}$ in eq(\ref{repwidth}) does not depend on the number density of electrons in each sources.
Thus, the ambiguity in the determination of the bandwidth is rather less than the ambiguity in the determination of
quantities depending on the electron number density.  

\vspace{0.1cm}
Secondly, 
The duration $t_W(\rm repeat)$ of the repeating FRBs is larger on average\cite{resmallflux, nonsmallflux, effelsberg121102} 
than the duration $t_W(\rm oneoff)$ of the one-off FRBs.
As we have explained in the section(\ref{L}), the duration of the repeating FRBs is in practice given by the time $\tau'_W\sim r_a/c$, 
while the duration of the one-off FRBs is given by the relaxation time $\tau$; within the time the energies of the 
oscillating electrons are thermalized.  The relaxation time caused by interactions among electrons 
is proportional to $1/T_c^2\propto 1/B^4$ since $T_c\propto B^2$.
Thus, the relaxation time of the one-off FRBs is much shorter at least by the several order of magnitude than that of the repeating FRBs
because the magnetic field of neutron star is stronger than that of accretion disk.
As we have explained above, the duration $\tau'_W\sim r_a/c$ of the repeating FRBs is shorter than the relaxation time $\tau(\rm repeating)$
of the FRBs; $\tau'_W<\tau(\rm repeating)$. Although we can not compare the duration $\tau'_W$ with the relaxation time $\tau(\rm oneoff)$,
it is naturally expected that  $\tau'_W>\tau(\rm oneoff)$.
Furthermore, the observed duration $t_W(\nu)$ of repeating FRB with frequency $\nu$, satisfies
$t_W(\nu)=\frac{\nu_0}{\nu} t_W(\nu_0)$. Then, in general we expect the inequality,

\begin{equation}
\mbox{Duration}; \quad \quad 
t_W(\nu,\mbox{repeating})=\frac{\nu_0}{\nu} t_W(\nu_0, \mbox{repeating})>\frac{\nu_0}{\nu}t_W(\nu_0,\rm oneoff)
\end{equation}

Indeed, If we see the data\cite{CHIME} of CHIME with the low frequency coverage $\nu=400$MHz $\sim 800$MHz,
their duration $t_W(\nu,\rm repeating)$ is 
much longer than the duration $t_W(1.4\rm GHz, oneoff)=2\sim 3$ms.
On the other hand, when we see the duration at much higher frequency $\nu>3$GHz than $1.4$GHz, 
the duration $t_W(\nu, \mbox{repeat})$ becomes smaller than the duration $t_W(1.4\rm GHz,\mbox{oneoff})$.

\vspace{0.1cm}
Thirdly, the luminosities of the one-off FRBs are larger on average\cite{resmallflux, nonsmallflux} than those of the repeating FRBs. 
It may be caused similarly by the difference of the strength of the magnetic field or Lorentz factor $\Pi(V,\theta)$,

\begin{equation}
\mbox{Luminosity}; \quad \quad \dot{W}(\mbox{oneoff})>\dot{W}(\mbox{repeat}) \quad \mbox{because}  \quad 
\dot{W}\propto \Pi(V,\theta)^3B^2n_e
\end{equation}
with  $\Pi(V,\theta)=\nu_c/(m_a/2\pi)$ for the repeating FRBs and $\Pi(V,\theta)\simeq 1$ for the one-off FRBs.
 
The stronger magnetic field of the neutron star leads to the larger luminosities of the bursts than those of the repeating FRBs.
The luminosities of the repeating FRBs decrease when the emitters of the bursts move to backward respect to the observer; $\Pi(V,\theta)<1$.
 Furthermore,
the number density of electrons emitting the bursts also affects their luminosities.
These causes would lead to the larger fluxes of  the one-off FRBs on average than the flux of the repeating FRBs.
If the effect of the Doppler factor is the most important, we can clearly observe the difference of the flux indicated by
the CHIME observation\cite{CHIME}. We should also stress that the ratio of $\dot{W}(\mbox{repeat})$ to $\dot{W}(\mbox{oneoff})$ depends on
the frequencies $\nu$ of the bursts compared respectively; $\dot{W}(\mbox{repeat})/\dot{W}(\mbox{oneoff})\propto (\nu/(m_a/2\pi))^3$.

\vspace{0.1cm}
It turns out that our model for the generation mechanism of the repeating and one-off FRBs can well explain the observed spectra-temporal features.
In the next section, we explain the new features of the FRBs recently observed based on the model.

\vspace{0.1cm}

\section{consistency of axion star model with recent observations}
\label{6}

We make several comments on recent observations, which have been published after our previous study\cite{narrowband}.
We have already confirmed the consistency of the axion star model with previous observations. 

\vspace{0.1cm}
First, 
a few or several sub-bursts have been observed\cite{drifting} within a burst in the repeating FRBs, 
while bursts composed of a single burst have been also observed. 
We speculate that 
the bursts involving sub-bursts are caused by the collisions between several axion stars closely packed with each other and the accretion disk,
while the single burst is caused by a single axion star collision with the disk.
The axion stars closely packed with each other are those like multiple star systems.
The density of the dark matter axions is especially high around galactic black hole 
so that the multiple axion star system or spatially compact clusters would be easily formed. Conceivably,
such closed packed axion stars could be formed by the fission of a massive unstable axion star\cite{axionstar} just like 
nuclear fission.  An unstable massive axion stars would dissociate to stable less massive axion stars.   
Thus, there are single axion stars or closed packed axion clusters around the galactic black hole.
They are the dark matter cloud and 
fall into black hole by gravitational perturbation.

\vspace{0.1cm}
Secondly,
the periodic activities\cite{periodicity12, periodicity18} have been observed in FRB121102 and FRB180916.J0158+65. 
The periods are $\sim157$ days and $\sim 16$ days,
respectively. These periodic activities can be understood in the following.
That is, there are some of axion stars which are alive after their collisions with the accretion disk.
They orbit black hole and periodically collide with the accretion disk. Such 
axion stars are those which collide the location of the disk with relatively weak magnetic field, for instance $eB=10^7$G.
They would be alive after their collisions with the accretion disk.
( For instance, when $M_a=10^{-11}M_{\odot}\sim 10^{43}$erg, $\dot{W}\sim 10^{40}$erg/s.  
If the thickness of the accretion disk is order of $1$km, the axion star pass it by $10^{-5}$second. 
It looses the energy $10^{-5}\dot{W}\sim 10^{35}$erg for the period. On the other hand, the length of the stretched axion star by tidal force is
of the order of $10^6$km, while the width is of the order of $0.1$km. 
The fraction of the mass $M_a$ passing the disk in the period is approximately $M_a(1\rm km/10^6km)\sim 10^{37}$erg. It is larger
than the energy dissipated in the period.
Thus, the axion star can survive in the collision with the accretion disk. )

Thus, the periodic activities could be caused by such axion stars
orbiting the black hole periodically. But eventually, they would be destroyed by the tidal force and the dissipation of their energies in the collisions.
As we have explained the cluster of the axion stars, they may form spatially compact cluster or widely spreaded cluster.
Such cluster of the axion stars would give rise to 
the active phase\cite{periodicity18} of the repeating FRB180916.J0158+65, 
in which spatially spreaded cluster of the axion stars collide the disk
periodically.

\vspace{0.1cm}
Recently, FRB200428\cite{coincidence,coincidence2} has been observed to be associated with magnetor SGR 1935+2154. 
Radio and X ray bursts have been detected simultaneously\cite{diffecoincidence}. Although the flux of the radio burst is $30$ times less than 
the weakest extragalactic FRB among those observed,
it strongly suggests that young magnetor is an origin of one-off FRBs.
But there are some observations\cite{nomagnetor} against the interpretation.
The precisely identified locations of the FRBs lie in the outskirts of host galaxies. 
In our model we interpret this coincidence of the radio burst and the X ray burst in the following.
That is, an axion star collision with the magnetor SGR 1935+2154 has induced the X ray emission as well as FRB200428.
The FRB is emitted by the electron gas in the surface of the magnetor. 
The collision suddenly deposits large amount energy to the
surface of the magnetor. The rapid energy injection would generate a quake in the surface or
lead to reconnection of strong magnetic field. It could cause
the X ray burst from the magnetor. Indeed, the arrival time of the radio wave is slightly earlier\cite{diffecoincidence} 
than the arrival time of X ray. Namely, there are two peaks of the radio bursts slightly prior to two X ray bursts, respectively. 
Two peaks are separated by the time $\sim 30$ms. 
In both peaks,
the peaks of the radio burst appear slightly earlier $\sim 5$ms than those of the X ray bursts.
It suggests that the axion star collision causes the X ray burst.

The smallness of the FRB200428 flux suggests that the axion star simply passes the neighborhood of the magnetor, not collides it.
Thus, the number density $n_e$ of electrons and the magnetic field $B$ in the neighborhood
is much smaller than the typical values on the surface of the magnetor.
We note the dependence of the luminosity on the magnetic field $\dot{W}\propto n_e B^2$.  

When we estimate the rate of the axion star collision with neutron star, we assume that
the radius $R_n$ of the neutron star is $10$km. Then, we have obtained the typical mass $M_a=10^{-12}M_{\odot}\sim 10^{-11}M_{\odot}$
of the axion star. We remember that the radius $R_a\propto M_a^{-1}$ and that the collision rate depends on 
both $R_a$ and $R_n$.
If the mass $M_a$ is smaller than these values, the rate of the collision is higher than the 
event rate of FRBs $10^{-3}$ per year in a galaxy.
Furthermore, when it passes the neighborhood of the neutron star, the rate of the 
passing is much larger than the even rate of FRBs. The passing the neighborhood implies large effective $R_a$
of the axion star in the collision.
Probably both effects of the small axion star mass and passing the neighborhood
causes the FRB200428 with small luminosity. 
Thus, the observation of the FRB200428 in our galaxy is not accidental.

\vspace{0.1cm}
Finally, most of FRBs show linear polarization, especially, the repeating FRBs show a large fraction of the linear polarization.   
The polarization angle 
of some bursts from the repeating FRBs 
do swing or some of them do not swing\cite{reswingpalarization,polarization}. 
On the other hand, the polarization angle of the one-off FRBs swings\cite{nonswingpolarization,swingpolarization}.
The swing of the polarization angle indicates that the radiation passes through the plasma
with magnetic field rapidly changing its direction.

These observations can be understood with the axion star model in the following.
The one-off FRBs arise from the surface of neutron star so that the radiations pass through
the magnetosphere.
The field configuration of the magnetosphere is magnetic dipolar so that the radiation passes through the region
with magnetic field rapidly changing the field configuration along the line of sight.
Such a swing of the polarization angle is also observed in pulsars.
On the other hand, the repeating FRBs arise from the magnetized accretion disk around galactic black hole.
The field configuration would be more complex than that of neutron star.
Thus, there would be a variety of polarization angle; some of them swing or do not swing.
Therefore, our generation mechanism of the FRBs is consistent with the observations of the polarization angle swing.

\section{determination of axion mass}
\label{7}
We would like to make a comment of the determination of the axion mass by using the spectrum of the FRBs.
In the previous paper we have estimated the axion mass $\sim 7\times 10^{-6}$eV by using the spectrum of the repeating FRB121102.
The spectrum of the burst B$33$ in the reference\cite{axionmass} shows the center frequency $\sim 1400$MHz with small bandwidth $\sim 50$MHz. 
The small bandwidth $\Delta\nu_{obs} \propto B$ indicates the small magnetic field $B$. Such a small magnetic field
would be present in a location far away from the black hole with mass $M_b$.  
We speculate that the gravitational red shift $z_g=(1/\sqrt{1-2GM_b/r}-1)$ is small. Similarly, 
the velocity of the location in the accretion disk would be
non relativistic. It implies that the Doppler effect by the velocity is small, i.e. $\Pi(V,\theta)\simeq 1$. Thus,
the radiation is affected by only cosmological red shift $z_c$. Because of these reasons,
we have estimated the axion mass

\begin{equation}
m_a=2\pi\nu_{obs}\Pi(V,\theta)^{-1}(1+z_c)(1+z_g)\simeq
\frac{\nu_{obs}}{1.4\mbox{GHz}}(1+z_c)(6\times 10^{-6}\mbox{eV})\simeq 7.2\times 10^{-6}\mbox{eV}, 
\end{equation}
with $z_c\simeq 0.2$, $\Pi(V,\theta)\simeq 1$ and $z_g \ll 1$.

In the formula, $\nu_{obs}$ denotes the center frequency of the repeating FRB, e.g. $\nu_{obs}=1.4$GHz for the burst B$33$.
The recent much sensitive observation\cite{bandwidth2} of the repeating FRB190711 presents a burst 
with the center frequency $1360$MHz and the small bandwidth $\sim 65$MHz.
Because the cosmological red shift $z_c\simeq 0.52$ of the host galaxy is known, 
we estimate the axion mass $\simeq 8.9\times 10^{-6}$eV
under the assumption that the Doppler effect of the accretion disk and gravitational red shift are neglected. 
Obviously these values represent lower limit of the real axion mass 
because we have neglected the effects of the gravitational red shift $z_g>0$
and Doppler effect $\Pi(V,\theta)<1$.  
We speculate that the real axion mass would be about $10^{-5}$eV.

On the other hand, the one-off FRBs is not affected by the Doppler effect due to the motion of the source.
The rotational velocity of the surface in neutron star is non relativistic.
Thus,
the Doppler factor is negligible; $\Pi(V,\theta)\simeq 1$. 
It implies that there are no variety of the center frequencies.
The center frequency of the burst is unique; $\nu_c=m_a/2\pi$,
although the line spectrum is thermal broadening.
The observed center frequency $\nu_{obs,c}$ of the one-off FRBs receives cosmological $z_c$ and gravitational $z_g$ red shift, 
$\nu_{obs,c}=(m_a/2\pi)/\big((1+z_c)(1+z_g)\big)$
where $z_g=(1/\sqrt{1-2GM_n/R_n}-1)$ with neutron star mass $M_n$ and radius $R_n$.
Therefore, we can determine the axion mass $m_a$ by observing the center frequency $\nu_{obs}$ of 
the one-off FRBs, if we know the cosmological and gravitational red shift,

\begin{equation}
m_a=2\pi\nu_{obs}(1+z_c)(1+z_g)=8.6\times 10^{-6}\mbox{eV}\frac{\nu_{obs}}{2\mbox{GHz}}(1+z_c)(1+z_g) \quad \mbox{for one-off FRB},
\end{equation}
where $\nu_{obs}$ represents the center frequency of the thermally broaden spectrum in the one-off FRBs,
although the center frequency has not yet been identified.

The cosmological red shift can be speculated by dispersion measure, or it can be precisely determined by the 
observing host galaxy.
On the other hand, the gravitational
red shift is difficult to estimate. The mass of neutron star has been observed to be in the range $1.5M_{\odot}<M_n<2M_{\odot}$, 
but the radius $R_n$ has not yet done. It is estimated by using models for nuclear matter such that $8\mbox{km}<R_n<12$km.
The ambiguity in the determination of $z_g$ is large.
However,
using both data on the repeating FRBs and one-off FRBs, 
we will be able to approximately obtain the axion mass with the above formulae.

\section{summary and discussion}
\label{8}

\vspace{0.3cm}
Our model for generation of FRBs is that repeating ( one-off ) FRBs are produced by the axion star collision with
accretion disk of galactic black hole with mass $10^3M_{\odot}\sim 10^5M_{\odot}$ ( neutron star ).
( The presence of two different types of the FRBs is supported by recent analysis\cite{evolution,different} 
of the number density evolution of non-repeating and repeating FRBs towards the early Universe. )
Based on the generation mechanism proposed in our previous papers\cite{axionmodel,narrowband},
we have explained recently observed spectral-temporal features of both repeating and one-off FRBs.
Especially, we have focused on the repeating FRBs, noting that the features of
the bursts are fairly affected by Doppler effect owing to the relativistic motion of accretion disk.
In particular, we have obtained the relation\cite{drift-duration,superradiance} 
observationally holding well between the downward drifting rate $\delta \nu_{obs}/\delta t_D$
and the duration $t_W$ in the repeating FRBs; $\delta \nu_{obs}/\delta t_D=-\nu_{obs} A/t_W$.
The relation has been pointed out in original papers\cite{drift-duration,superradiance} 
in which superradiance from molecules or atoms has been proposed as
a generation mechanism of the repeating FRBs. 
The arguments in the papers are very general. Applying the arguments to our generation mechanism, 
we have derived the relation. 
The essence of the relation is Doppler effect caused by the accretion disk, which
moves with relativistic velocity in neighborhood of the black hole. 

According to our model, all of the FRBs have line spectra thermally broadened. Among them,   
the repeating FRBs have narrow bandwidth $\Delta\nu\propto \nu_c$ ( $50\mbox{MHz}<\Delta\nu <1\mbox{GHz}$ ).
The relativistic motion of the source ( accretion disk )
causes several specific features of the repeating FRBs. 

First, their center frequencies $\nu_c$ take
various values $\nu_c$ ( $200\mbox{MHz}<\nu_c <8\mbox{GHz}$ ). 
The various center frequencies arise from the various velocities ( due to differential rotation ) of the accretion disk.
The radio waves are emitted by the accretion disk moving toward or backward to the observer. The center frequency is
Doppler shifted such that
$\nu_c=\Pi(V,\theta)m_a/2\pi$ with Doppler factor $\Pi(V,\theta)=\sqrt{1-V^2}/(1-V\cos\theta)$. 

The harmonically oscillating electrons emit the radio waves with the line spectrum.
But the line spectrum is thermally broadened; $\Delta\nu=\nu_c\sqrt{T_c/m_e}$. 
The temperature
$T_c$ is determined by magnetic field of the source ( accretion disk and neutron star ). 
Using the observed approximate relation $\Delta\nu\sim 0.16\nu_c$ of repeating FRBs, 
we have derived the strength of the magnetic field $eB\sim 10^{10}$G in
the accretion disk, assuming $M_a=4\times 10^{-12}M_{\odot}$ and $m_a=0.6\times 10^{-5}$eV.

Secondly, the duration $t_W(\mbox{repeat})$ of the repeating FRBs is larger on average than duration $t_W(\mbox{oneoff})$ of the one-off FRBs. 
In particular, the tendency is more obvious when we compare the duration of the burst at the lower frequency such as $1$GHz or less.
The duration of the repeating FRBs with, for instance, the frequency $\sim 600$MHz is shifted to longer duration 
than those of the FRBs with center frequency $\sim 1.5$GHz because of  
accretion disk moving backward to the observer. On the other hand, the one-off FRBs are not affected by
such Doppler effect because the rotation velocity of neutron star is negligibly small.
Therefore, the duration of the repeating FRBs with the frequency $\sim 1$GHz or less
is longer than those of the one-off FRBs; $t_W(\mbox{repeat})>t_W(\mbox{oneoff})$.

Additionally, the flux densities of the repeating FRBs are smaller on average than that of the one-off FRBs.
Especially, the tendency is more obvious for the flux densities of the bursts with lower frequencies.
This is because they are more strongly affected by the Doppler factor $\Pi(V,\theta)$. 

\vspace{0.1cm}
The strength of the magnetic field $eB\sim 10^{10}$G estimated in the accretion disk 
is a hundred times smaller than the typical magnetic field $\sim10^{12}$G of neutron stars.
The difference makes the bandwidths $\Delta\nu\propto eB$ of the one-off FRBs a hundred times larger than those of
the repeating FRBs. Thus, the bandwidths of the one-off FRBs are over the extent of receiver frequency.
We predict that the bursts with their spectra showing upper or lower limit in frequency are repeating FRBs when they are
observed in the range of the frequency $400$MHz to $4$GHz.

\vspace{0.1cm}
Furthermore, we can explain the periodicity in the repeating FRB121102 and FRB180916.J0158+65.
The periodicity is caused by the periodic orbits of the axion stars. Some of the axion stars can survive 
the collision with thin accretion disk so that they may collide periodically.

The association of FRB200428 with magnetor SGR 1935+2154 is interpreted as the 
axion star collision with the magnetor. The axion star collision with the magnetor stimulates X ray emission.
Thus, the burst of FRB200428 arrives slightly earlier than the X ray burst, as it has been observed.
We interpret it as a consequence of the axion star passing the neighborhood of the magnetor or the axion star colliding 
with smaller mass than $10^{-12}M_{\odot}$. 
Then, we understand the smallness of the luminosity of FRB200428 and the reason why we can observe it in our galaxy;
the rate of the collision is higher than the event rate of FRBs with standard luminosity.

Furthermore, some of repeating FRBs show the polarization angle swing, while some do not show the swing.
On the other hand, the one-off FRBs show the polarization angle swings.
The phenomena arise from the presence of the magnetosphere around neutron star or complex configuration of magnetic field in the accretion disk.

\vspace{0.1cm}

QCD axion is a hypothetical particle and its presence has been extensively explored\cite{ADMX}. 
As we have shown above, once we admit the existence, the mystery of the FRBs is clearly solved. 
In other words, the success of the axion model for the FRBs suggests the reality of the axion. 
But we need to confirm the existence. In order to do so, we have proposed\cite{newmethod} a new method for detecting the QCD axion 
with the use of superconductor and existing radio telescope ( or sensitive radio receiver ). 
The type 2 superconductor under magnetic field emits dipole radiations
converted from the dark matter axion. The emission mechanism is identical to the one for the FRBs.
The dark matter axion induces oscillating electric field under magnetic field imposed on the superconductor.
Because the magnetic field penetrates only the surface of the superconductor,
Cooper pairs in the surface emit the radiations owing to the oscillating electric field. 
The flux from the surface area $S$ of the superconductor, e.g. Mb$_3$Sn is 
estimated to be $\sim 10^{-18}$W$\big(S/60\mbox{cm}^2\big)^2\big(B/3\times 10^4 \rm G\big)^2$ 
under the magnetic field $5\times 10^4$G.
The frequency of the radiations is given by $m_a/2\pi$. 
We expect the frequency ranging $1.5$GHz to $2.5$GHz.
The corresponding axion mass is $6\times 10^{-6}$eV to $10^{-5}$eV.
It would be easy to detect the radio wave with such a large flux 
by existing radio telescope or sensitive radio receiver.

.

\vspace{0.2cm}
The author
expresses thanks to Yasuhiro Kishimoto, Izumi Tsutsui and Osamu Morimatsu for useful comments  
and discussions.

.

\end{document}